\documentclass[12pt]{elsarticle}

\usepackage{newtxtext,newtxmath}
\usepackage[utf8x]{inputenc}
\usepackage[english]{babel}
\usepackage{amsmath,amsbsy,amsfonts}
\usepackage{graphicx}
\usepackage{subfig}
\usepackage{color}
\usepackage{float}
\usepackage[numbers]{natbib}
\usepackage[left=1cm,right=1cm,top=2cm,bottom=2cm]{geometry}
\usepackage{physics}
\usepackage[colorlinks=true, allcolors=red]{hyperref}
\usepackage{subcaption}
\numberwithin{equation}{section}

\begin{document}

\begin{frontmatter}

\title{Exact Dynamics and Bloch Oscillations in a Non-Hermitian Zigzag Glauber–Fock Lattice} 

\author[label0,label1]{G.S. Fahara-Ojeda} 
\author[label1]{L.A. Tapia-Alarcón} 
\author[label1,label2]{B.M. Villegas-Martínez} 
\author[label2]{G.K. Thiruvathukal}
\author[label3]{A. Francisco Neto}
\author[label4]{S. Gesing} 
\author[label5]{T. Battelle}
\author[label6]{V. Stankovski}
\author[label7]{I.A. Bocanegra-Garay}
\author[label8]{A. Venegas-Gomez}
\author[label1]{J.J. Escobedo-Alatorre}
\author[label1]{R.A. Beltrán-Vargas} 

\affiliation[label0]{organization={Macke Scholar, Johannes Kepler University Linz},
addressline={Altenbergerstrasse 69, A-4040 Linz}, 
country={Austria.}}

\affiliation[label1]{organization={Centro de Investigación en Ingeniería y Ciencias Aplicadas (CIICAp),Universidad Autónoma del Estado de Morelos},
            addressline={Av. Universidad 1001}, 
            city={Cuernavaca},
            postcode={62209}, 
            state={Morelos},
            country={Mexico.}}

\affiliation[label2]{organization={ Loyola University Chicago},
            city={Chicago},
            state={Illinois},
            country={USA.}}       

\affiliation[label3]{organization={DEPRO Escola de Minas UFOP},
            addressline={Campus Morro do Cruzeiro}, 
            city={Ouro Preto},
            postcode={35400-000}, 
            state={Minas Gerais},
            country={Brazil.}}

\affiliation[label4]{organization={San Diego Supercomputer Center, University of California},
            addressline={9500 Gilman Drive, La Jolla}, 
            city={San Diego},
            postcode={92093-0505}, 
            state={California},
            country={USA.}} 

\affiliation[label5]{organization={Arizona State University}, 
            state={Arizona},
            country={USA.}}         
            
\affiliation[label6]{
    organization={Faculty of Computer and Information, Science University of Ljubljana},  
    state={Ljubljana},
    country={Slovenia.}}

\affiliation[label7]{
    organization={Departamento de Física Teórica, Atómica y Óptica, and Laboratory for Disruptive Interdisciplinary Science, Universidad de Valladolid}, 
    postcode={47011}, 
    state={Valladolid},
    country={Spain.}}

    \affiliation[label8]{
    organization={QURECA Ltd., Glasgow, G2 4JR},  
    state={Scotland},
    country={United Kingdom.}}

\begin{abstract}
The discrete dynamics of a one-dimensional non-hermitian zigzag waveguide array is studied theoretically. This system is characterized by unbalanced left/right nearest-neighbor hopping and reciprocal second-neighbor interactions in both hopping directions.  The interplay between the unbalanced hopping amplitudes and the waveguide setup results in Non-Hermitian Bloch oscillations, where energy transport is either amplified or dissipated at specific locations, depending on the relative strengths of the left and right hopping terms.  We demonstrate that by applying an appropriate non-unitary transformation, the original system can be mapped into an equivalent reciprocal waveguide with Hermitian dynamics. This transformation allows us to derive a closed-form analytic solution, which is then compared with the numerical solution of the system.
\end{abstract}


\end{frontmatter}

\section{Introduction}

Non-Hermitian photonics has rapidly emerged as a new frontier in physics and engineering, unveiling novel pathways to control and guide light propagation beyond traditional Hermitian frameworks \cite{1,2,3}. The foundations of this field trace back to the seminal theoretical contributions of Bender et al. \cite{4,5,6,7}, who demonstrated that certain non-Hermitian Hamiltonians could exhibit entirely real spectra when the PT symmetry is satisfied. This groundbreaking achievement set the stage for the manipulation of photonic eigenstates by incorporating elements such as optical gain-loss and non-reciprocal interactions, thereby laying the cornerstone for the field. Since then, a large number of theoretical and experimental efforts have led to the discovery of a variety of exciting effects, such as unidirectional invisibility \cite{8,9}, fast and slow light phenomena near exceptional points \cite{10,11,12}, and complex Bloch oscillations \cite{13,14}, among others \cite{15,16,17}.

A prominent platform for exploring non-Hermitian physics in photonics is the use of photonic lattices (arrays of coupled optical waveguides) due to their ability to precisely manipulate asymmetric and non-reciprocal propagation of light \cite{18,19,20,21,22,23,24,25,26,27}. 
Typically, non-Hermiticity in these systems is introduced via two primary mechanisms: spatially distributed optical gain (amplification) and loss (absorption), widely used in PT-symmetric designs \cite{22,25,27,28,29,30}, and asymmetric inter-site coupling, inspired by the Hatano-Nelson model, which governs nonreciprocal hopping in one-dimensional lattices \cite{31,32,33}. Although these lattices exhibit intriguing spectral and topological properties, obtaining closed-form solutions for studying their complex dynamics remains a significant challenge, especially under open boundary conditions \cite{34,35,36,37,38,39}. Naturally, several approaches have been developed to explore the underlying physics governing these systems. Among them, non-unitary transformations have proven to be a powerful tool, providing a systematic method to map non-Hermitian systems to their Hermitian counterparts \cite{40}. For instance, non-unitary transformations have been reported to enable precise control of dissipation and amplification terms in structured photonic systems \cite{36,41,42,43,44}. A notable application of this approach has been applied to the Glauber-Fock lattice \cite{45,46,47}, resulting in a non-Hermitian Hamiltonian that models an anisotropic waveguide array, akin to the Hatano-Nelson type \cite{48}. However, most prior studies have primarily focused on non-reciprocal nearest-neighbor hopping, often overlooking the role of second-nearest-neighbor interactions. This limitation restricts the exploration of more complex wave dynamics in realistic photonic setups, such as quantum computing and quantum information processing, where compact and efficient optical circuits demand a more comprehensive understanding of higher-order interactions \cite{49}.
Therefore, the contribution of this work lies in the study of a class of exactly solvable one-dimensional non-Hermitian zigzag waveguide system, a variant of the semi-infinite Glauber-Fock lattice. This configuration incorporates nonreciprocal nearest-neighbor hopping amplitudes while maintaining symmetric next-nearest-neighbor couplings, allowing for a broader exploration of non-Hermitian transport phenomena beyond conventional models \cite{34,35,36,37,38,39}. Notably, we show that the non-Hermitian Hamiltonian of the system can be mapped to its Hermitian counterpart by a non-unitary transformation —more general than those previously proposed \cite{48}—, enabling to obtain an exact analytical solution, which is subsequently validated by its numerical solution, thereby extending our earlier work in \cite{50}. Additionally, the zigzag configuration offers a rich mathematical structure and serves as a natural platform for exploring non-Hermitian spatial Bloch oscillations, an area that has been largely unexplored in non-Hermitian Glauber-Fock-type lattices. Our results reveal that by tuning the forward and backward hopping amplitudes between nearest-neighbor waveguides, it is possible to induce controlled amplification or attenuation in the Bloch oscillations.

\section{Main equation of the model}

\begin{figure}[H]
\centering
{\includegraphics[width=0.7 \textwidth]{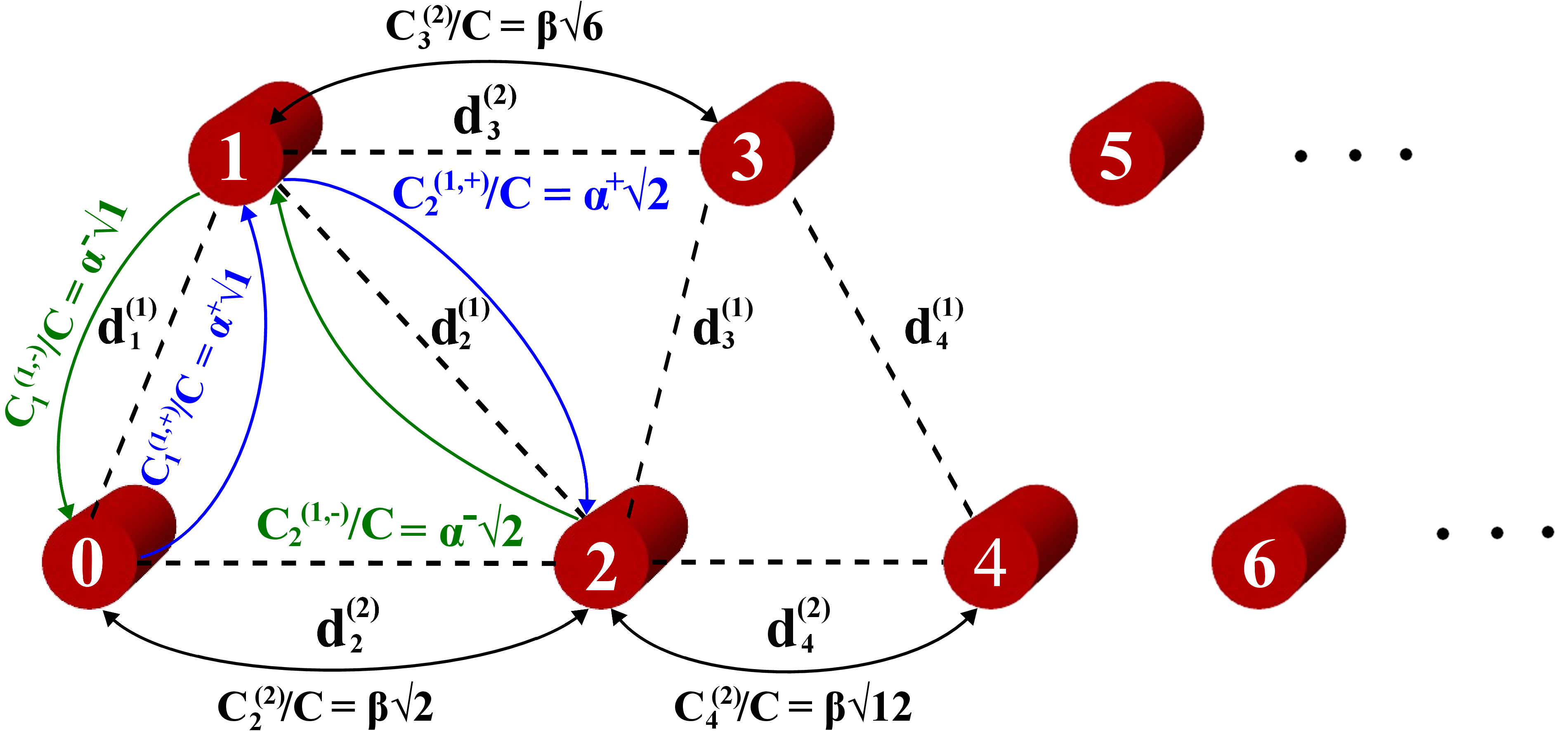}}
\caption{Schematic diagram of the zigzag waveguide system, with $\alpha^{\pm}$ representing the site-independent forward and backward hopping amplitudes between adjacent waveguides and $\beta$ the hopping almplitude for second interaction.}
\label{f1}
\end{figure}

The proposed photonic lattice model is a non-Hermitian zigzag array with two interleaved 1D single-mode waveguides in a scalene arrangement. The light dynamics in this generic lattice, with asymmetric couplings between adjacent sites and interactions extending up to the second order, is described by the following discrete set of coupled equations
\begin{equation} \label{1}
i \frac{d\mathcal{E}_n(z)}{dz}  + \mu_n \mathcal{E}_n(z) + C_n^{(1,-)}  \mathcal{E}_{n-1}(z) + C_{n+1}^{(1,+)} \mathcal{E}_{n+1}(z)  + C_{n}^{(2)} \mathcal{E}_{n-2}(z) + C_{n+2}^{(2)} \mathcal{E}_{n+2}(z) =0,\quad
n=0,1,2,...
\end{equation}
where $\mathcal{E}_{n}(z)$ represents the complex field amplitude along the propagation distance $z$ at the $n$-th site,  with the boundary condition $ \mathcal{E}_{n}(z)=0 $ for $n<0$.  Each waveguide has a propagation constant $\mu_{n}$. The setup of our model is depicted in Fig. \eqref{f1}, where adjacent waveguides in the neighboring layer are coupled using evanescent fields in the transverse direction. These couplings are defined by $C_n^{(1,-)}=\alpha^{-} C  \exp[-\frac{d^{(1)}_{n}-d_{1}}{\kappa}]$ and $C_{n+1}^{(1,+)}=\alpha^{+}  C \exp[-\frac{d^{(1)}_{n+1}-d_{1}}{\kappa}]$. Here, $\alpha^{\pm}$ are constants that describe the site-independent forward and backward hopping amplitudes between adjacent waveguides, $C$ is a reference coupling coefficient, $d^{(1)}_{n}=d_{1}-\frac{\kappa}{2} \ln(n)$ denotes the distance to the first-order neighbor of the $n$-th site, $d_{1}$ is a reference distance adjustable between the first two sites, and $\kappa$ is a free parameter determined by coupled mode theory \cite{51,52,53,54,55}. Furthermore, the coupling of the evanescent field tail from next-nearest neighbors in the same layer is characterized by $C^{(2)}_{n}=\beta C \exp[-\frac{d^{(2)}_{n}-d_{2}}{\kappa}]$ where $\beta$ is a fixed constant hopping amplitude. Here, $d^{(2)}_{n}=d_{2}-\frac{\kappa}{2} \ln[n \left(n-1\right)]$ represents the distance to the $n$-th site’s second-order neighbor. 

This model represents a modification of the semi-infinite Glauber-Fock lattice, characterized by hopping amplitudes that increase with the square root of the site number. As the site index $n$ increases, the waveguides progressively approach each other, intensifying the coupling between the adjacent waveguides. In our model, we assume that the propagation constant of the $n$-th waveguide is given by $\mu_{n}=\mu + \alpha_{0} n$, where $\alpha_{0}$ is a linear gradient factor. This gradient acts as an effective external potential that induces Bloch oscillations \cite{52,56,57,58,59,60}. In particular, $\alpha_{0} > 0$ implies that the refractive index of the waveguides gradually increases with the waveguide number $n$. Although $\alpha_{0}$ could be negative, leading to a different lattice response, we exclusively consider positive values for $\alpha_{0}$ in this study. Therefore, if we substitute $\mathcal{E}_{n}(z)=\Psi_{n}\left(Z\right) \exp\left(i \mu z\right)$, the Eq.~\eqref{1} can be expressed in a dimensionless form
\begin{equation} \label{2}
i \frac{d \Psi_{n}(Z)}{dZ} + \lambda n  \Psi_{n}(Z) + \alpha^{-} \sqrt{n} \Psi_{n-1}(Z) + \alpha^{+} \sqrt{n+1} \Psi_{n+1}(Z) 
+ \beta \left[\sqrt{n(n-1)} \Psi_{n-2}(Z) + \sqrt{(n+1)(n+2)} \Psi_{n+2}(Z) \right] =0, 
\end{equation} with $\lambda=\alpha_{0}/C$ and where $\Psi_{n}(Z)$ is a function of the scaled (or normalized) distance $Z=C z$. It should be noted in Eq.~\eqref{2} that setting $\alpha^{+}=\alpha^{-}=\alpha_{1}$ and $\beta=\alpha_{2}$ recovers the Hermitian Zigzag lattice reported in previous studies \cite{61}. However, in the non-Hermitian case discussed here, $\alpha^{+} \neq \alpha^{-}$.

Although the proposed system is initially presented as a theoretical model, its experimental realization could nowadays be feasible by using femtosecond laser-writing technology in polished fused silica, which enables the fabrication of waveguide systems with flexible geometries \cite{49,62}. In this setup, the coupling between neighboring waveguides can be controlled by adjusting the inscription depth and spacing, while propagation constants are tuned via thermo-optical or electro-optical effects \cite{56,63}. Non-Hermiticity, introduced by asymmetric left/right hopping amplitudes, can be implemented through photonic gauge fields \cite{64,65} or synthetic gauge fields via combined phase and amplitude modulations \cite{13,66,67}.

\section{Exact solution}

To obtain the most general form of the exact analytical solution for the case where $\alpha^{+} \neq \alpha^{-} \neq 0$ and $\beta \neq 0$, we use the method described in \cite{68}. Here, we adopt a simplified notation where each waveguide field is a component of the single state vector $\ket{\psi\left(Z\right)}=\sum_{n=0}^{\infty} \Psi_{n}(Z) \ket{n}$. In this representation, $\ket{n}$, which serves as the waveguide basis vector, is analogous to a Fock state and corresponds to the scenario where only the $n$-th site is excited. In this way, the system of differential equations given by Eq. \eqref{2} is equivalent to the following Schrödinger-like equation
\begin{equation} \label{3}
i\frac{d \ket{\psi\left(Z\right)}}{dZ} =\hat{H} \ket{\psi\left(Z\right)},
\end{equation}
where the corresponding Hamiltonian, expressed in terms of the bosonic annihilation, $\hat{a}|n\rangle = \sqrt{n}|n-1\rangle$, and creation, $\hat{a}^{\dagger}|n\rangle = \sqrt{n+1}|n+1\rangle$, operators is given by 
\begin{equation} \label{4}
\hat{H}=- \left[ \lambda \hat{a}^{\dagger} \hat{a} + \alpha^{-} \hat{a}^{\dagger} + \alpha^{+} \hat{a}  + \beta \left(\hat{a}^{\dagger2}  + \hat{a}^2\right)  \right].
\end{equation}
The corresponding Hamiltonian can be rewritten using the representation of the generators $\hat{K}^{+}=\frac{\hat{a}^{\dagger2}}{2}$, $ \hat{K}^{-}=\frac{\hat{a}^{2}}{2}$ and $ \hat{K}^{0}=\frac{\left(\hat{a}^{\dagger} \hat{a} + 1/2\right)}{2}$, fulfilling the Lie algebra su(1,1) commutation relations $[\hat{K}^{0}, \hat{K}^{\pm}] = \pm \hat{K}^{\pm}$, \quad $[\hat{K}^{+}, \hat{K}^{-}] = -2\hat{K}^{0}$, and also satisfying the following commutation relations with the one-photon operators $[\hat{K}^{+}, \hat{a}] = -\hat{a}^{\dagger}, \quad [\hat{K}^{-}, \hat{a}^{\dagger}] = \hat{a}, \quad \left[\hat{K}^{0}, \hat{a}^{\dagger} \right]=\frac{\hat{a}^{\dagger}}{2}, \quad \left[\hat{K}^{0}, \hat{a} \right]=-\frac{\hat{a}}{2}.$ Notice two key aspects of the non-Hermitian Hamiltonian inside the brackets. First, it is non-\(\mathcal{PT}\)-symmetric, meaning it is not invariant under the combined operations of space-time inversion: \(\hat{\mathcal P}:\hat{a}\rightarrow -\hat{a}\) and the time-reversal operator \(\hat{\mathcal T}: \hat{a}\rightarrow \hat{a}\), with similar transformations for \(\hat{a}^{\dagger}\) \cite{69,70,71}. This lack of symmetry arises due to the presence of linear terms in \(\hat{a}\) and \(\hat{a}^{\dagger}\). Second, to simplify the analysis and obtain a physically meaningful representation, we aim to apply a change of variable that includes a non-unitary transformation to remove these terms. This process should transform the non-conservative system into a conservative one, represented by a new Schrödinger-like equation with a Hermitian Hamiltonian, making the system algebraically more manageable. In this context, due to the algebraic structure of the non-Hermitian Hamiltonian, we define the change of variable variable $\ket{\psi\left(Z\right)}= \hat{U}^{-1}_{1} \ket{\phi\left(Z\right)}$ in Eq. \eqref{3}, where the non-unitary transformation is given by
\begin{equation}  \label{5}
\hat{U}_{1}=\exp\left( \zeta^{-} \hat{a}^{\dagger} -\zeta^{+} \hat{a} \right),
\end{equation} 
with $\zeta^{+}$ and $\zeta^{-}$ being two parameters to be determined. This transformation is analogous to the Glauber displacement operator \cite{72}. Upon applying it along with the previous commutation relations and the use of the formula $e^{\hat{A}}\hat{B}e^{-\hat{A}}=\hat{B}+ \left[\hat{A},\hat{B}\right]+ \frac{1}{2!}\left[\hat{A}, \left[\hat{A},\hat{B}\right]\right]+\frac{1}{3!}\left[\hat{A}, \left[\hat{A}, \left[\hat{A},\hat{B}\right]\right]\right]+ \ldots$ \cite{73,74}, we arrive at the transformed Schrödinger-like equation 
\begin{align}  \label{6}
i\frac{d \ket{\phi\left(Z\right)}}{dZ} &=\Bigg\lbrace -2 \beta \left( \hat{K}^{+} + \frac{\lambda}{\beta} \hat{K}^{0} + \hat{K}^{-}  \right) + \hat{a}^{\dagger} \left(  \zeta^{-} \lambda + 2 \beta \zeta^{+} -\alpha^{-} \right) +\hat{a} \left( \zeta^{+} \lambda + 2 \beta \zeta^{-} -\alpha^{+} \right)\Bigg\rbrace  \ket{\phi\left(Z\right)} \nonumber \\
& + \Bigg\lbrace \zeta^{-} \alpha^{+}  + \zeta^{+} \alpha^{-} - \lambda \left( \zeta^{+} \zeta^{-} - 1/2  \right) - \beta \left( \zeta^{-2} + \zeta^{+2} \right)    \Bigg\rbrace  \ket{\phi\left(Z\right)}.
\end{align} 
To get rid of the non-Hermitian linear terms and obtain a Hermitian propagation description, we impose the following conditions:
\begin{align}  \label{7}
\zeta^{-} \lambda + 2 \beta \zeta^{+} -\alpha^{-}&=0, \nonumber \\
\zeta^{+} \lambda  + 2 \beta \zeta^{-} -\alpha^{+}&=0.
\end{align} 
Solving this system yields
\begin{equation}  \label{8}
\zeta^{\mp}=\frac{2\beta \alpha^{\pm} -\lambda \alpha^{\mp}}{\Gamma^{2}},
\end{equation}  
with $\Gamma=\sqrt{4 \beta^{2} - \lambda^{2}}$, which solution holds as long as $\lambda \neq -2 \beta$. The case $\lambda= -2 \beta$ requires a different straightforward approach, not covered here. Substituting Eq. \eqref{8} into Eq. \eqref{6}, one gets
\begin{equation}  \label{9}
i\frac{d \ket{\phi\left(Z\right)}}{dZ} =\left[ -2 \beta \left( \hat{K}^{+} + \frac{\lambda}{\beta} \hat{K}^{0} + \hat{K}^{-}  \right) + f \right] \ket{\phi\left(Z\right)}, 
\end{equation} 
with $f=\frac{\lambda}{2}-\frac{\alpha^{+} \alpha^{-} }{ \Gamma^{2}} \left[ \lambda-\beta\left(\frac{\alpha^{-}}{\alpha^{+}} + \frac{\alpha^{+}}{\alpha^{-}} \right) \right]$.
Note that the Hermitian dynamics in the modified Schrödinger-like equation resembles a squeezed-like lattice when identifying $\beta \rightarrow{1}$ and $\lambda\rightarrow{\alpha}$ \cite{75}. Alternatively, it admits an equivalent Hermitian representation analogous to the photon production mechanism in the dynamical Casimir effect under the threshold off-resonance condition, where $\lambda$ acts as a frequency shift \cite{76}. Thus, the present waveguide system and the associated non-unitary transformation could be effectively employed to emulate these physical phenomena.

It is also insightful to contrast this transformation with the previously introduced non-unitary transformation in Ref. \cite{48}. If we implement the non-unitary transformation $\hat{U}_{1}=\exp\left( -\hat{n} \upsilon \right)$, where $\upsilon$ is a parameter to be determined, it allows the factorization of the coefficients of $\hat{a}$ and $\hat{a}^{\dagger}$ into a common factor, thus eliminating non-Hermiticity in linear terms of the Hamiltonian \eqref{4}. However, in this approach, non-Hermiticity is transferred to the quadratic terms.  Consequently, the non-unitary transformation proposed in \cite{48} is particularly effective when $\beta=0$, which is not the case in this work. Indeed, applying our non-unitary transformation to the Hatano–Nelson-type system of \cite{48} would produce Hermitian dynamics in which only the free propagation term remains.

The formal solution of Eq. \eqref{9} is
\begin{equation}  \label{10}
\ket{\phi\left(Z\right)}=\exp\left(-i f Z \right) \exp\left[2 i \beta \left( \hat{K}^{+} + \frac{\lambda}{\beta} \hat{K}^{0} + \hat{K}^{-}  \right) Z \right] \ket{\phi\left(0\right)}, 
\end{equation} 
and the original solution for $\ket{\psi\left(Z\right)}$ is recovered by using the inverse transformation $\ket{\phi\left(Z\right)}= \hat{U}_{1} \ket{\psi\left(Z\right)}$, to obtain
\begin{equation}  \label{11}
\ket{\psi\left(Z\right)}=\exp\left(-i f Z \right) \exp \left[-\left( \zeta^{-} \hat{a}^{\dagger} -\zeta^{+} \hat{a} \right) \right] \exp\left[2 i \beta \left( \hat{K}^{+} + \frac{\lambda}{\beta} \hat{K}^{0} + \hat{K}^{-}  \right) Z \right] \exp\left( \zeta^{-} \hat{a}^{\dagger} -\zeta^{+} \hat{a} \right) \ket{\psi\left(0\right)}.
\end{equation} 
In this case, we have
\[
\exp\left[-2 i \beta \left( \hat{K}^{+} + \frac{\lambda}{\beta} \hat{K}^{0} + \hat{K}^{-}  \right) Z \right] \left( \zeta^{-} \hat{a}^{\dagger} -\zeta^{+} \hat{a} \right) \exp\left[2 i \beta \left( \hat{K}^{+} + \frac{\lambda}{\beta} \hat{K}^{0} + \hat{K}^{-}  \right) Z \right]= \hat{a}^{\dagger} \left(\zeta^{-}- \xi^{+} \right)-\hat{a} \left( \zeta^{+} -\xi^{-}  \right),\] with $\xi^{\pm} \left(Z\right) = \frac{\alpha^{\mp}}{\Gamma^{2}} \left[ 2 \left( \lambda - 2 \beta \frac{\alpha^{\pm}}{\alpha^{\mp}} \right)  \sinh^{2} \left( \Gamma Z/2 \right) \pm i \Gamma \sinh \left( \Gamma Z \right)  \right]$. 
Following the approach in \cite{61} and applying the identity $\exp\left[\frac{1}{2} \left( \varepsilon_{1} \varepsilon_{2} +  2 \varepsilon_{2} \eta_{1} + \eta_{2} \eta_{1} \right) \right] \exp\left( \varepsilon_{1} \hat{a}^{\dagger} -\varepsilon_{2} \hat{a} \right) \exp\left( \eta_{1} \hat{a}^{\dagger} -\eta_{2} \hat{a} \right)=   \exp\left[ \left( \varepsilon_{1} + \eta_{1} \right) \hat{a}^{\dagger} \right] \exp\left[ -\left( \varepsilon_{2} + \eta_{2} \right) \hat{a} \right]$, we can express the evolution operator as a product of six exponentials, up to a global phase factor $\exp\left[-i\nu \left(Z\right) \right]$. This decomposition enables to evaluate the action of each exponential operator on an arbitrary initial state, which can be explicitly written as follows 
\begin{align} \label{12}
\ket{\psi\left(Z\right)}& =
\exp\left[-i\nu \left(Z\right) \right] \exp\left[-\frac{\xi^{+}\left(Z\right) \xi^{-}\left(Z\right) }{2} \right]
\exp\left[g_{1}(Z) \hat{K}^{+}\right] \exp\left[g_{0}(Z) \hat{K}^{0}\right] \exp\left[g_{1}(Z) \hat{K}^{-}\right] \nonumber \\
& \times \exp\left[\xi^{+} \left(Z\right) \hat{a}^{\dagger} \right]  \exp\left[-\xi^{-} \left(Z\right) \hat{a}\right]\ket{\psi\left(0\right)},
\end{align}
being
\begin{equation} 
\begin{aligned} \label{13}
\nu(Z)=f Z - \frac{\zeta^{+} \zeta^{-}}{\Gamma} \left[ \lambda + \beta \left( \frac{\zeta^{-}}{\zeta^{+}} + \frac{\zeta^{+}}{\zeta^{-}} \right)  \right] \sinh\left( \Gamma Z \right),\quad 
g_{0}(Z)=&-2 \ln \left[\cosh(\Gamma Z)-i \frac{ \lambda}{\Gamma} \sinh(\Gamma Z)\right],\\
g_{1}(Z)=\frac{2i \beta \sinh\left(\Gamma Z\right)}{\Gamma \cosh\left(\Gamma Z\right) - i \lambda  \sinh\left(\Gamma Z\right)}.&
\end{aligned}
\end{equation}
The factorization of the exponential operator involving the su(1,1) generators is performed using the Omega Matrix Calculus \cite{77}, as shown in Appendix A.  If we excite the waveguide array by injecting a light beam
into the $n$th waveguide —thereby setting the initial state as $\ket{\psi\left(0\right)}=\ket{n}$—, the light amplitude in the $m$-th waveguide at position $Z$, can be obtained as $\Psi^{\left(n\right)}_{m}\left(Z \right)=\braket{m}{\psi(Z)}$. Following the steps detailed in Appendix B of \cite{61} (see also Appendix A), we obtain a closed-form analytical expression describing the propagation dynamics along the waveguide array
\begin{equation}\label{14}
\Psi^{\left(n\right)}_{m}\left(Z \right)=\exp\left[-i\nu \left(Z\right) \right]
\sum_{k=0}^{\infty}\mathscr{S}_{m,k}\left(Z\right) \mathfrak{D}_{k,n}\left(Z\right),
\end{equation}
where
\begin{align} \label{15}
& \mathscr{S}_{m,k}\left(Z\right) 
= \mathscr{Q}_{m,k}\left(Z\right) \sum_{p=0}^{\infty} \frac{ 2^p \Theta\left(m-p \right)  \Theta\left(k-p \right) 
\cos^2\left[\left(m-p \right)\frac{\pi}{2}  \right] \cos^2\left[\left(k-p \right)\frac{\pi}{2}  \right] \exp\left[\frac{g_0 \left(Z\right)}{2} p \right]}{g_1^p\left(Z\right) \left(\frac{m-p}{2} \right) ! \left(\frac{k-p}{2} \right) ! p!},  \nonumber \\
& \mathscr{Q}_{m,k}\left(Z\right) = \sqrt{m!k!}\exp\left[\frac{g_0\left(Z\right)}{4} \right] \left[\frac{g_1\left(Z\right)}{2} \right] ^{\frac{m+k}{2}}, \\
& \mathfrak{D}_{k,n}\left(Z\right) =  \exp\left[-\frac{\xi^{+}\left(Z\right) \xi^{-}\left(Z\right) }{2} \right] \nonumber
\begin{cases}
\sqrt{\frac{n!}{k!}} \left[\xi^{+}\left(Z\right)\right]^{k-n} L_n^{\left(k-n \right) }\left[ \xi^{+}\left(Z\right) \xi^{-}\left(Z\right)\right], & k\geq n,
\\
\sqrt{\frac{k!}{n!}}\left[-\xi^{-}\left(Z\right)\right]^{n-k} L_k^{\left(n-k \right) } \left[ \xi^{+}\left(Z\right) \xi^{-}\left(Z\right) \right], & k<n,
\end{cases}
\end{align}
with the step function $\Theta(x)= \begin{cases} 0, & x<0,\\ 1, & x \geq 0 \end{cases}$ and $L_\mathscr{r}^{(\mathscr{l})}\left(x\right)$ representing the associated Laguerre polynomials of order $\mathscr{r}$.

\section{Results}

In this section, the predictions of our theoretical analysis are validated through the direct numerical integration of Eq. \eqref{2},  which is solved using a high-precision variable step Runge-Kutta-Fehlberg method.  To illustrate a physically realistic scenario, we consider parameter values reported experimentally that would support the design of a photonic lattice composed of approximately 40 coupled waveguides. To the best of our knowledge, current femtosecond laser writing techniques allow the fabrication of up to 60 single-mode waveguides in fused silica, with parameters compatible with those of the Glauber–Fock photonic lattice \cite{48}, as demonstrated in \cite{62}. 

We examine two distinct regimes based on asymmetry in the back-and-forth hopping amplitudes. In the case where $\alpha^{-}>\alpha^{+}$,  the coupling coefficients between the zeroth and first waveguides are set to $C_1^{(1,-)}=0.88 cm^{-1}$ and $C_1^{(1,+)}=0.792 cm^{-1}$, corresponding to a reference coupling strength of $C=0.44 cm^{-1}$, with hopping amplitudes $\alpha^{-}=2$ and $\alpha^{+}=1.8$. In contrast, for the case $\alpha^{-}<\alpha^{+}$, we consider $C_1^{(1,-)}=0.792 cm^{-1}$ and $C_1^{(1,+)}=0.88 cm^{-1}$ being $\alpha^{-}=1.8$ and $\alpha^{+}=2$. These values are in accordance with the Hermitian case, where $C_1=C_1^{(1,+)}=C_1^{(1,-)}=0.88 cm^{-1}$, as reported experimentally \cite{78}. For both configurations, we fix the linear gradient to $\alpha_{0}=0.044 mm^{-1}$ \cite{79}, resulting in $\lambda=1$, and set the second-order hopping amplitude to $\beta=0.15$. Under these conditions, where $\lambda \gg 2\beta$, the quantity $\Gamma$ becomes purely imaginary, and the system's hyperbolic dynamics Eq. \eqref{13} transitions into a trigonometric oscillation. This leads to spatial Bloch oscillations, characterized by a period $Z_{p}=\frac{2\pi}{\left| \Gamma \right|}=\frac{2\pi}{\sqrt{\lambda^2-4\beta^2}}\approx 6.58$ \cite{61}, ensuring the observation of at least one full Bloch-like oscillation cycle within the simulated propagation range of $Z = 10$.

\begin{figure}[H]
    \centering
	\subfloat[Numerical simulation of Eq.~\eqref{2} using the Runge-Kutta-Fehlberg method.]
	{\includegraphics[width=0.42\textwidth]{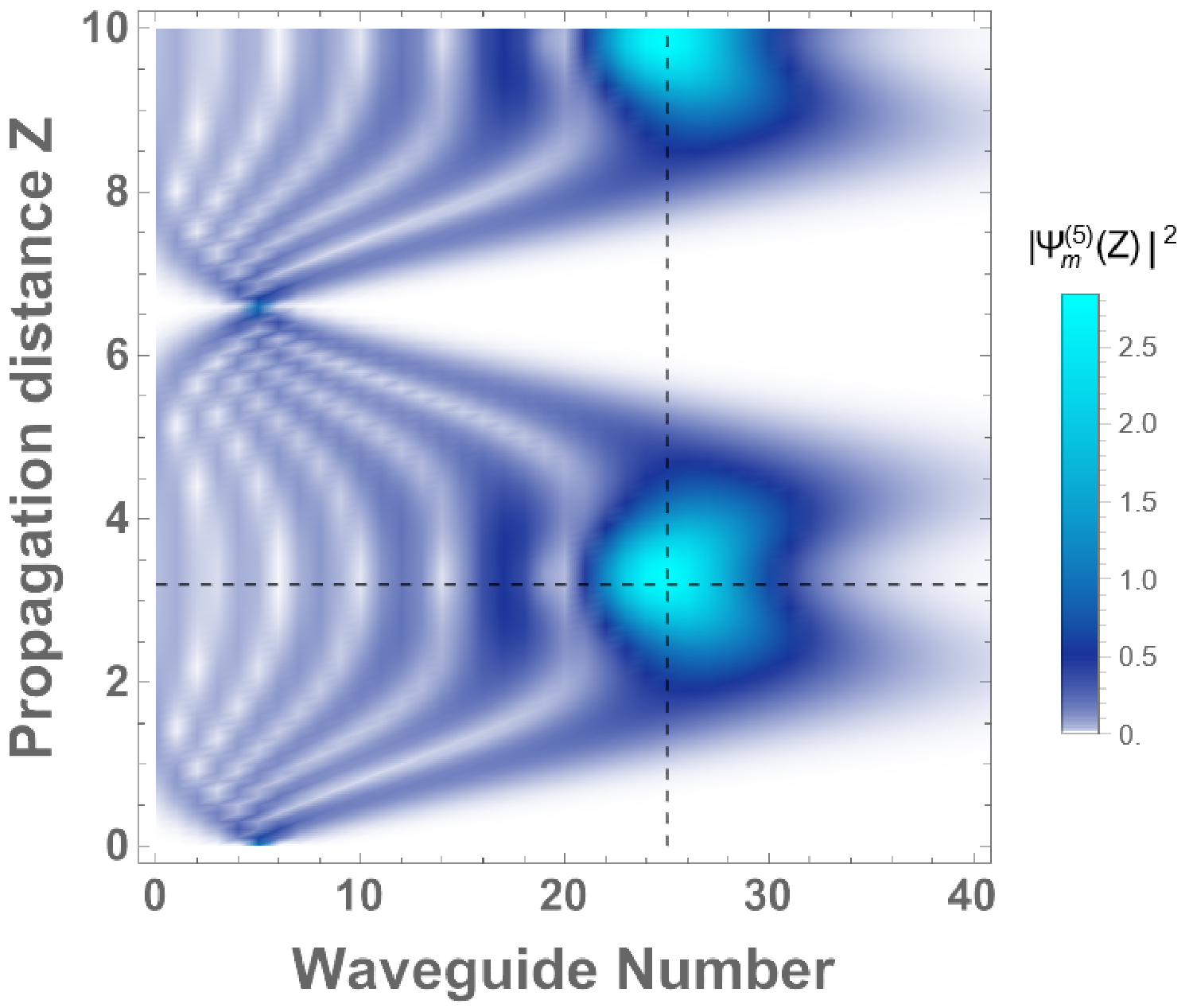}}
	\qquad
	\subfloat[Theoretical predictions derived from Eq.~\eqref{14}.]
	{\includegraphics[width=0.42\textwidth]{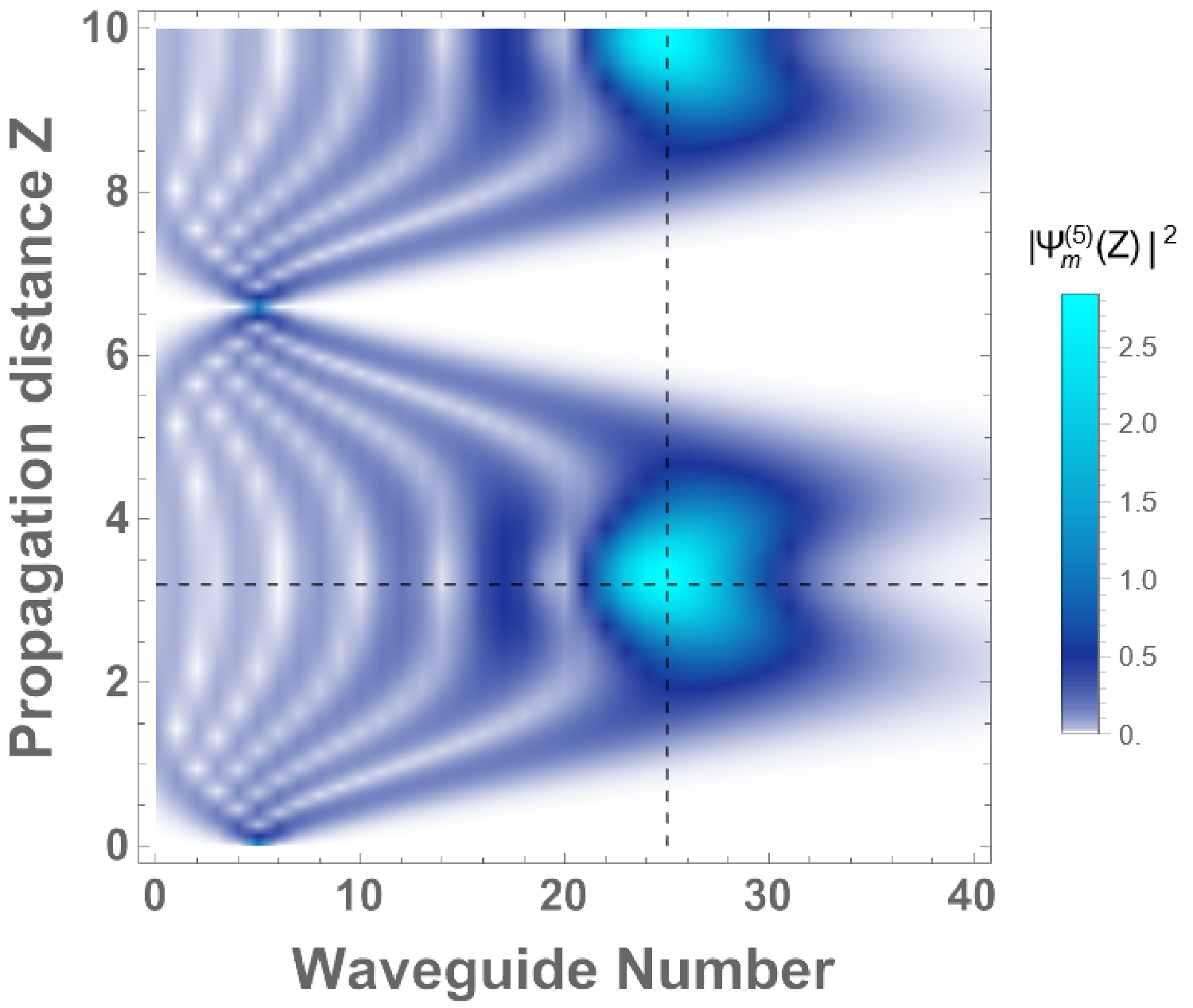}} \\
	\subfloat[Exact versus numerical solutions at $m=25$]
	{\includegraphics[width=0.42\textwidth]{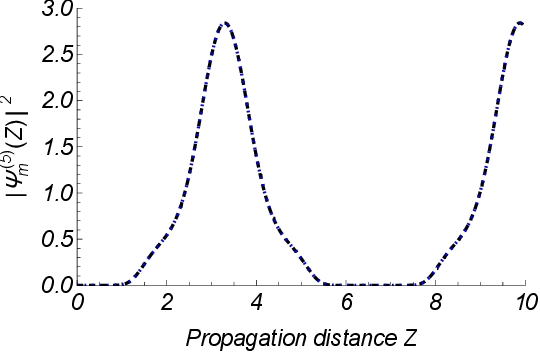}}
	\qquad
	\subfloat[Exact versus numerical solutions at $Z=3.2$]
	{\includegraphics[width=0.42\textwidth]{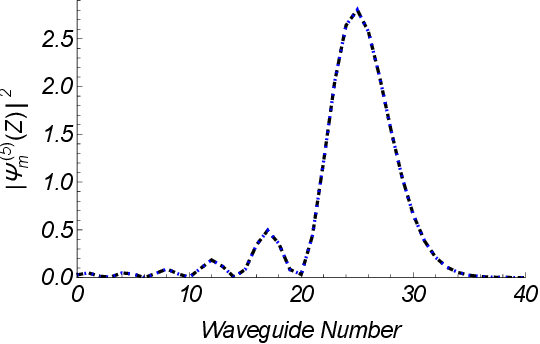}}
	\caption{Light intensity propagation, $I^{(5)}_m = |\Psi^{(5)}_m(Z)|^2$, for the case $\alpha^{-} > \alpha^{+}$, with light initially launched at site $n = 5$. Panel (a) shows the exact analytical solution, while panel (b) displays the corresponding numerical simulation. Both illustrate the evolution of the optical wavepacket along the propagation direction $Z$, highlighting the emergence of non-Hermitian Bloch-like oscillations and asymmetric energy transport due to the unbalanced coupling. Panels (c) and (d) provide complementary views to quantify the system's dynamics. In panel (c), the light intensity is plotted at a fixed waveguide site, $m = 25$, as a function of propagation distance $Z$, revealing a clear oscillatory behavior with optical amplification. Panel (d) shows the spatial intensity profile across the lattice at a fixed propagation distance, $Z = 3.2$, where the asymmetric transport becomes visible as the wavepacket shifts preferentially toward one side of the array. In both panels (c) and (d), the black dashed lines correspond to the numerical simulation, while the blue circles denote the exact analytical results, showing good agreement between both solutions. All plots were generated using $n = 5$, $\lambda = 1$, $\alpha^{+} = 1.8$, $\alpha^{-} = 2$, and $\beta = 0.15$.}
	\label{f2}
\end{figure}

\begin{figure}[H]
    \centering
	\subfloat[Numerical simulation of Eq.~\eqref{2} using the Runge-Kutta-Fehlberg method.]
	{\includegraphics[width=0.42\textwidth]{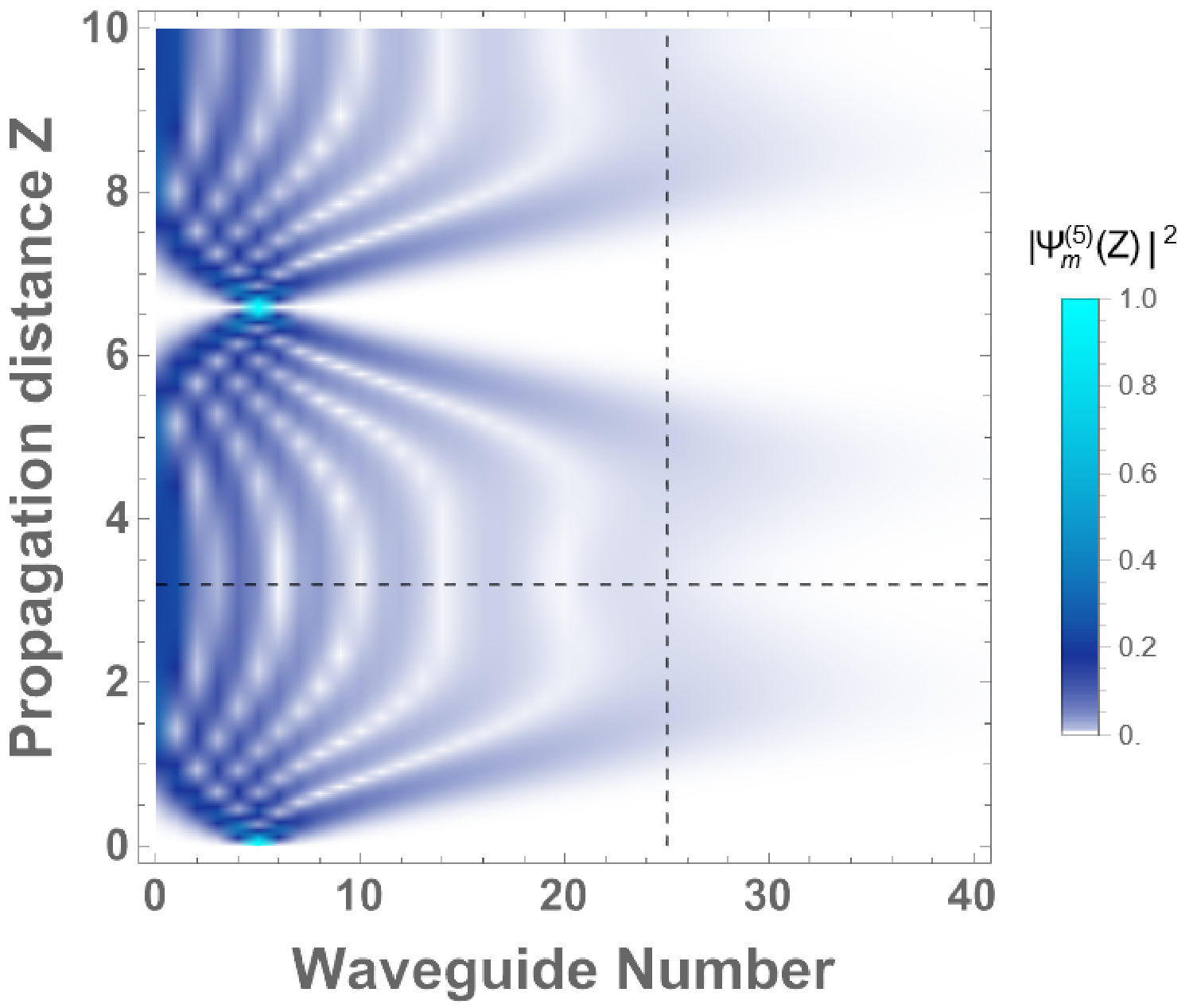}}
	\qquad
	\subfloat[Theoretical predictions derived from Eq.~\eqref{14}.]
	{\includegraphics[width=0.42\textwidth]{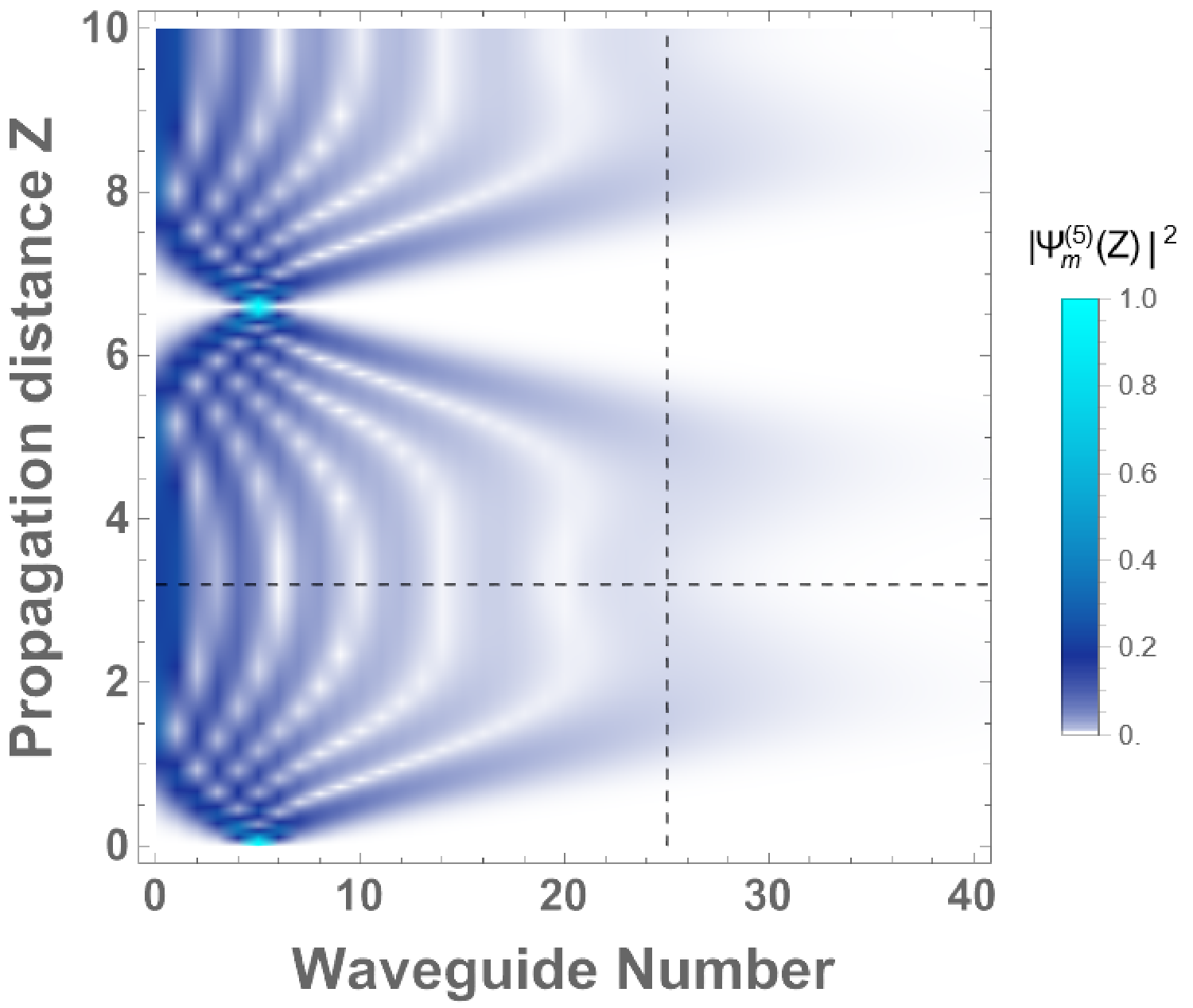}} \\
	\subfloat[Exact versus numerical solutions at $m=25$]
	{\includegraphics[width=0.42\textwidth]{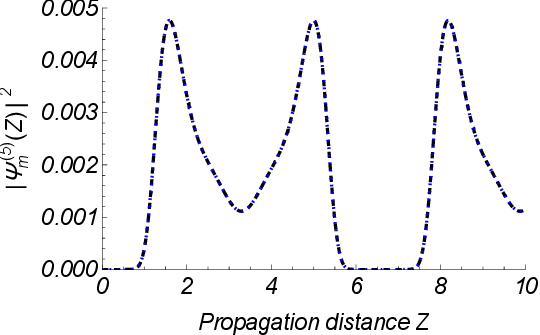}}
	\qquad
	\subfloat[Exact versus numerical solutions at $Z=3.2$]
	{\includegraphics[width=0.42\textwidth]{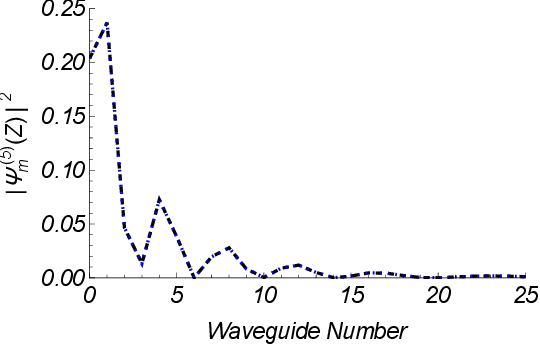}}
	\caption{Light intensity propagation, $I^{(5)}_m = |\Psi^{(5)}_m(Z)|^2$, for the case $\alpha^{+} > \alpha^{-}$, with light initially launched at site $n = 5$. Panel (a) shows the exact analytical solution, while panel (b) displays the corresponding numerical simulation. In this regime, both figures show the overall light intensity diminishes, yet the wave packet continues to undergo Bloch-like oscillations with a pronounced decay in amplitude. Panels (c) and (d) provide complementary views to the intensity at a fixed waveguide site, $m = 25$, as a function of propagation distance $Z$, the curve displays Bloch-like oscillations with a clear, decaying envelope, indicating that the overall intensity diminishes as the wave packet propagate while panel (d) presents the spatial intensity profile at a fixed propagation distance, $Z = 3.2$, where a pronounced asymmetric attenuation is observed, with light intensity significantly reduced on one side of the array. In both panels (c) and (d), the black dashed lines represent the numerical simulation, and the blue circles denote the exact analytical results, showing good agreement between both solutions. All plots were generated using $n = 5$, $\lambda = 1$, $\alpha^{+} = 2$, $\alpha^{-} = 1.8$, and $\beta = 0.15$.}
	\label{f3}
\end{figure}

\begin{figure}[H]
    \centering
	\subfloat[$\alpha^{-} > \alpha^{+}$.]
	{\includegraphics[width=0.42\textwidth]{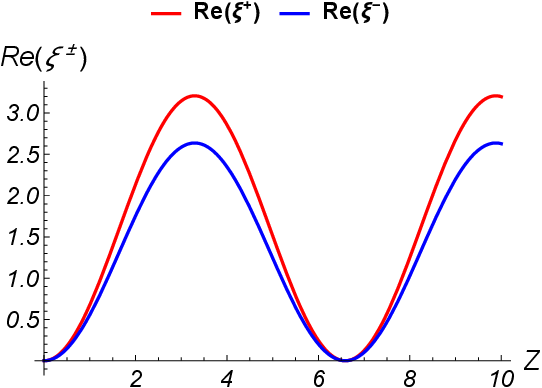}}
	\qquad
	\subfloat[$\alpha^{-} > \alpha^{+}$.]
	{\includegraphics[width=0.42\textwidth]{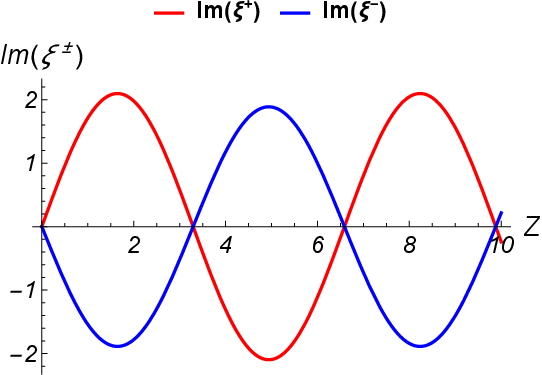}} \\
	\subfloat[$\alpha^{+} > \alpha^{+}$.]
	{\includegraphics[width=0.42\textwidth]{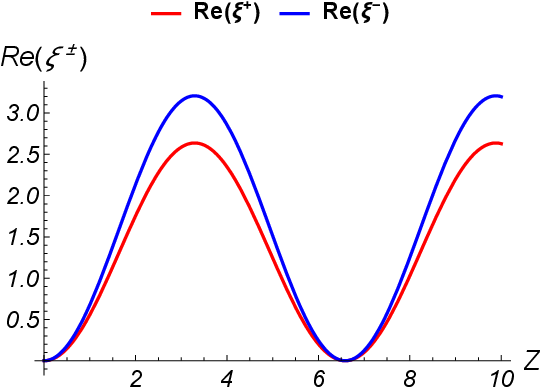}}
	\qquad
	\subfloat[$\alpha^{+} > \alpha^{+}$.]
	{\includegraphics[width=0.42\textwidth]{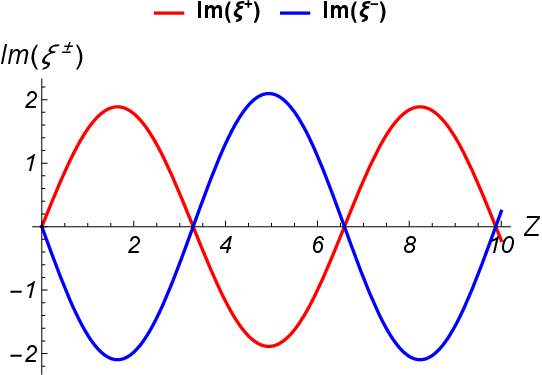}}
	\caption{Comparison between the real (a, c) and imaginary (b, d) parts of $\xi^{+}(Z)$ and $\xi^{-}(Z)$ for the two non-reciprocal hopping regimes,  $\alpha^{-} > \alpha^{+}$ and $\alpha^{+} > \alpha^{-}$. These complex functions dictate whether the propagating light wave packet undergoes amplification or attenuation within the non-Hermitian zigzag lattice.  In the case of $\alpha^{-} > \alpha^{+}$, $\xi^{+}(Z)$ (red curves) dominates both the real and imaginary components comparate to $\xi^{-}(Z)$  (blue curves), leading to amplification. Conversely, when $\alpha^{+} > \alpha^{-}$, $\xi^{-}(Z)$ prevails, inducing attenuation. This interplay between $\xi^{+}(Z)$ and $\xi^{-}(Z)$ underpins the non-reciprocal light transport in the non-Hermitian Bloch-like oscillations.}
	\label{f4}
\end{figure}

Figure \eqref{f2} shows the distribution of the light intensity, $I^{\left(5\right)}_{m}=\abs{\Psi^{\left(5\right)}_{m}\left(Z \right)}^2$, when light is injected into the site $n = 5$ for the case $\alpha^{-}>\alpha^{+}$. Two distinct behaviors emerge: one where the light spreads across the array and another where it localizes,  indicative of Bloch-like oscillations. Unlike their Hermitian counterparts, which exhibit a symmetric energy distribution profile, the non-Hermitian dynamics induce strongly asymmetric transport. The light-wave packet exhibits a pronounced directional bias, preferentially propagating toward one side of the lattice. Figs. \eqref{f2}(a) and \eqref{f2}(b), respectively show numerical and exact solutions for the wave packet spectrum evolution, clearly demonstrating Bloch-like oscillations with exponential amplitude growth and significant optical power amplification during propagation. Figs. \eqref{f2}(c) and \eqref{f2}(d) further depict the light intensity profiles as a function of the propagation distance at waveguide site $m = 25$, and as a function of waveguide number at fixed propagation distance $Z = 3.2$, respectively. In both cases, the blue circles—which mark the theoretical predictions derived from our exact analytical solution Eq. \eqref{14}—are in excellent agreement with the numerical simulations of light dynamics governed by Eq. \eqref{2}.
In the regime $\alpha^{+}>\alpha^{-}$, where the forward hopping amplitude exceeds the backward amplitude,  the overall light intensity decreases compared to the $\alpha^{-}>\alpha^{+}$ case shown in Fig. \eqref{f2}. However, both the exact analytic solution and numerical simulations in Figs. \eqref{f3} (a) and \eqref{f3} (b) confirm that the wave packet continues to undergo Bloch-like oscillations, albeit with a pronounced decay in amplitude. As the wave packet propagates, its intensity steadily decreases, highlighting the overall attenuation. Panels (c) and (d) further illustrate this effect by plotting the intensity at waveguide $m=25$ and the distribution at $Z=3.2$. Both graphs show a clear trend toward localization and rapid attenuation. In all cases, the exact analytical solution matches the numerical results. These two regimes can be understood in terms of the mathematical structure of the complex functions $\xi^{+}(Z)$ and $\xi^{-}(Z)$ coming from the Glauber-like displacement operator in Eq. \eqref{12}, whose imaginary parts determine whether the propagation leads to amplification or attenuation. Specifically, when $\alpha^{+} > \alpha^{-}$, the contribution from $\xi^{-}(Z)$ predominates,  leading to attenuation in the system; conversely, when $\alpha^{-} > \alpha^{+}$, the dynamics are governed by $\xi^{+}(Z)$, which provides amplification. Figure \eqref{f4} illustrates this behavior by plotting the real and imaginary parts of $\xi^{+}(Z)$ and $\xi^{-}(Z)$ for both regimes, clearly highlighting the previously-mentioned effects. In the Hermitian case where $\alpha= \alpha^{+}=\alpha^{-}$, the contributions from amplification and attenuation balance each other, leading to symmetric propagation.

Although asymmetric transport in non-Hermitian systems is often linked to the non-Hermitian skin effect \cite{80} — a phenomenon where a macroscopic number of eigenstates accumulate at the boundaries — the behavior observed here is different. In our system, directional localization and asymmetric intensity profiles emerge from the interplay between the gain and loss processes encoded in $\xi^{+}(Z)$ and $\xi^{-}(Z)$,  as well as from the lattice setup. This finding is consistent with previous studies \cite{1,13, 31} showing that non-reciprocal hopping can induce directional energy transport without necessarily invoking the conventional skin effect.

\section{Conclusions}

We have presented an analytically solvable model of a one-dimensional non-Hermitian zigzag Glauber–Fock waveguide lattice, with nonreciprocal nearest-neighbor hopping and symmetric next-nearest-neighbor couplings. Using a single non-unitary transformation, we demonstrated that the non-Hermitian zigzag waveguide array can be mapped into an equivalent Hermitian optical lattice. This transformation naturally induces Hermitian dynamics akin to a squeezed-like lattice, allowing thus to obtain closed-form solution for light dynamics applicable to different initial conditions. As a result, the propagation in the non-conservative (non-Hermitian) and conservative (Hermitian) dynamics is linked via this non-unitary transformation, enabling an easy transition between these two descriptions. The exact analytical solution not only fits the numerical simulations but also establishes a robust framework for understanding and controlling light within this waveguide configuration. 

We showed that, by carefully selecting parameters, non-Hermitian Bloch-like oscillations with a pronounced directional bias in wave packet transport can emerge. In this context, the complex functions $\xi^{+}(Z)$ and $\xi^{-}(Z)$ —which naturally arise from applying the inverse non-unitary transformation to analyze the dynamics of non-Hermitian propagation— provide insight into the control of attenuation and amplification processes in the array. Directional attenuation, primarily driven by  $\xi^{-}(Z)$  in the regime $\alpha^{+}>\alpha^{-}$ leads to localized optical fields where losses outweigh amplification. Conversely,  when $\alpha^{-}>\alpha^{+}$, $\xi^{+}(Z)$  contributes to the light amplification. By tuning the imbalance between forward and backward hopping amplitudes, we can control optical amplification or attenuation during the Bloch oscillations. 

This ability to manage directional transport and tunable amplification suggests promising applications in quantum computing, where the zigzag lattice model could serve as a photonic platform for new approaches to quantum state preparation and manipulation via adiabatic processes in a non-hermitian scenario. For example, it could useful to analyzed cases where the waveguide array is injected with a path-entangled state and where multiple guides are illuminated under the amplification regime—modeling external electromagnetic input to mitigate losses. Moreover, the interplay between the zigzag configuration and nearest-neighbor interactions could modify the Bloch oscillation period for any input state, yielding a spatial oscillation period consistent with previous reports in the Hermitian case \cite{61}. Finally, a freely available graphical user interface for reproducing the numerical results has been published on Zenodo \cite{81}, enabling users to replicate the findings presented in this article.

\section{Acknowledgment}

L.A. Tapia-Alarcón acknowledges SECIHTI for the financial support through his master's scholarship (CVU 1317353). The work of I. A. B.-G. is supported by Spanish MCIN with funding from European Union Next Generation EU (PRTRC17.I1) and Consejeria de Educacion from Junta de Castilla y Leon through QCAYLE project, as well as Grant No. PID2023-148409NB-I00 MTM funded by AEI/10.13039/501100011033, and RED2022-134301-T. Financial support of the Department of Education, Junta de Castilla y Le\'on, and FEDER Funds is also gratefully acknowledged (Reference: CLU-2023-1-05).

\section{Ethics declarations}
\subsection{Ethics Approval}
This material is the author's original work, which has not been previously published elsewhere. All authors have been personally and actively involved in the substantial work that led to the paper and will take public responsibility for its content.

\subsection{Conflict of Interest}
The authors declare that they have no competing interests in the publication of this manuscript.

\appendix 
\section{Disentangling Formula Using the Omega Matrix Calculus}

The Omega Matrix Calculus (OMC) is a combinatorial method that extends to matrix analysis MacMahon's partition analysis originally devised to study the partition of natural numbers in pure mathematics (see, e.g., \cite{77} and references therein). Here we illustrate its utility by obtaining the disentangling formula for the exponential containing generators of the su(1,1) Lie algebra in Eq. \eqref{12} and the matrix elements in the oscillator realization of the su(1,1) Lie algebra leading to $\mathscr{S}_{m,k}\left(Z\right)$. Let
\begin{align}\label{XEnt}
\hat{X}(\boldsymbol{A})
=\exp\left(A_+\hat{K}^++A_0\hat{K}^0+A_-\hat{K}^-\right)
\end{align} where $\boldsymbol{A}=(A_+,A_0,A_-)$ and $\hat{K}^{\pm,0}$ are the generators of the su(1,1) Lie algebra.
We want to obtain a formula for $f$, $g$, and $h$ in terms of $\alpha_{\pm,0}$ such that
\begin{align}\label{XDisent}
\hat{X}(\boldsymbol{A})
=\exp\left(f\hat{K}^+\right)\exp\left(g\hat{K}^0\right)\exp\left(h\hat{K}^-\right)
\end{align} in an alternative way that does not require solving a system of coupled ODE's. In order to do so we use the $2\times 2$ matrix representation of su(1,1) namely (see, e.g., \cite{82})
\[
\hat{K}^+\equiv 
\left(\begin{array}{cc}
0&1\\
0&0
\end{array}\right),\quad
\hat{K}^0\equiv 
\frac{1}{2}\left(\begin{array}{cc}
1&0\\
0&-1
\end{array}\right),\quad {\rm and}\quad
\hat{K}^-\equiv 
\left(\begin{array}{cc}
0&0\\
-1&0
\end{array}\right)
\] along with the OMC \cite{77}. The Omega operator
\[
\overset{\xi}{\underset{=
}{\Omega}}
\] acts on convergent matrix-valued functions depending on $\xi \in \mathbb{C}$ taken near the unit circle centered at the origin of the complex plane and extracts only terms with the powers of $\xi^n$ such that $n=0$. This process is often called the elimination of the Omega variable. For example, we have
\begin{align*}
\overset{\xi}{\underset{=
}{\Omega}}\frac{1}{(1-a\xi)(1-b/\xi)}
=\overset{\xi}{\underset{=
}{\Omega}}(1+a\xi+a^2\xi^2+\cdots)
\left(1+\frac{b}{\xi}+\frac{b^2}{\xi^2}+\cdots\right)=1+ab+a^2b^2+\cdots=\frac{1}{1-ab}
\end{align*} since the terms containing powers of $\xi$ present are all of the type $a^m\xi^m(b^n/\xi^n)$ and the action of the Omega operator selects only those terms such that $m=n$.

Let $z\in \mathbb{C}$ be small such that the Neumann series 
\begin{align}\label{Neu}
\left(\hat{I}-\frac{z\hat{A}}{\xi}\right)^{-1}=\sum_{n\geq 0}\frac{z^n\hat{A}^n}{\xi^n}
\end{align} is defined for any $\hat{A}$. Then the following identity holds
\begin{align}\label{Omegaexp}
e^{\hat{A}}=\overset{\xi}{\underset{=
}{\Omega}}e^{\frac{\xi}{z}}\left(\hat{I}-\frac{z\hat{A}}{\xi}\right)^{-1}.
\end{align} This result is stated in \cite[Lemma~2.3]{77}, but for completeness we reproduce the proof here. Indeed, using the Neumann series in Eq. \eqref{Neu} we have
\begin{align*}
\overset{\xi}{\underset{=
}{\Omega}}e^{\frac{\xi}{z}}\left(\hat{I}-\frac{z\hat{A}}{\xi}\right)^{-1}
=\overset{\xi}{\underset{=
}{\Omega}}\left(1+\frac{\xi}{1!z}+\frac{\xi^2}{2!z^2}+\cdots\right)
\left(\hat{I}+\frac{z\hat{A}}{\xi}+\frac{z^2\hat{A}^2}{\xi^2}+\cdots\right)
=\hat{I}+\frac{\hat{A}}{1!}+\frac{\hat{A}^2}{2!}+\cdots
=e^{\hat{A}}.
\end{align*} From now on we omit the complex variable $z$, but implicitly use it to ensure convergence of the Neumann series in Eq. \eqref{Neu}.

We are now ready to give an OMC based proof of Eq. \eqref{12}. We first consider Eq. \eqref{XEnt} using Eq. \eqref{Omegaexp}. Note that Eq. \eqref{Omegaexp} is particularly useful in this case since we need only the inverse of a $2\times 2$ matrix to eliminate the Omega variable and there is no need to compute the Jordan canonical form of $\hat{X}(\boldsymbol{A})$ in Eq. \eqref{XEnt}. This observation also holds in general whenever a matrix representation is available, and this comprises our main motivation of drawing the reader's attention to OMC in order to disentangle $\hat{X}(\boldsymbol{A})$. We have
\begin{align}\label{X}
\hat{X}(\boldsymbol{A})
&=\exp\left(A_+\hat{K}^++A_0\hat{K}^0+A_-\hat{K}^-\right)\nonumber\\
&=\exp\left(\begin{array}{cc}
A_0/2&A_+\\
-A_-&-A_0/2
\end{array}\right)\nonumber\\
&=\overset{\xi}{\underset{=
}{\Omega}}e^{\xi}\left(\begin{array}{cc}
1-A_0/(2\xi)&-A_+/\xi\\
A_-/\xi&1+A_0/(2\xi)
\end{array}\right)^{-1}\nonumber\\
&=\overset{\xi}{\underset{=
}{\Omega}}\frac{e^{\xi}}{1-\frac{A}{4\xi^2}}\left(\begin{array}{cc}
1+A_0/(2\xi)&A_+/\xi\\
-A_-/\xi&1-A_0/(2\xi)
\end{array}\right),
\end{align} where $A\equiv A_0^2-4A_+A_-$. It is easy to see that
\begin{align}\label{PFD}
\frac{1}{1-\frac{A}{4\xi^2}}
=\frac{1}{2}\left(\frac{1}{1-\frac{\sqrt{A}}{2\xi}}+\frac{1}{1+\frac{\sqrt{A}}{2\xi}}\right)
\end{align} using partial fraction decomposition. Going back to Eq. \eqref{X} and using the partial fraction decomposition in Eq. \eqref{PFD} we obtain
\begin{align*}
\overset{\xi}{\underset{=
}{\Omega}}\frac{e^{\xi}\left(1\pm \frac{A_0}{2\xi}\right)}{1-\frac{A}{4\xi^2}}
&=\frac{1}{2}\left(\overset{\xi}{\underset{=
}{\Omega}}\frac{e^{\xi}\left(1\pm \frac{A_0}{2\xi}\right)}{1-\frac{\sqrt{A}}{2\xi}}+\overset{\xi}{\underset{=
}{\Omega}}\frac{e^{\xi}\left(1\pm \frac{A_0}{2\xi}\right)}{1+\frac{\sqrt{A}}{2\xi}}\right)\\
&=\frac{1}{2}\left(e^{\sqrt{A}/2}\left(1\pm \frac{A_0}{\sqrt{A}}\right)+e^{-\sqrt{A}/2}\left(1\mp \frac{A_0}{\sqrt{A}}\right)\right)\\
&=\cosh\left(\frac{\sqrt{A}}{2}\right)\pm \frac{A_0}{\sqrt{A}}\sinh\left(\frac{\sqrt{A}}{2}\right)
\end{align*} and
\begin{align*}
\overset{\xi}{\underset{=
}{\Omega}}\frac{e^{\xi}A_{\pm}}{\xi\left(1-\frac{A}{4\xi^2}\right)}=\frac{A_{\pm}}{2}\left(\overset{\xi}{\underset{=
}{\Omega}}\frac{e^{\xi}}{\xi\left(1-\frac{\sqrt{A}}{2\xi}\right)}+\overset{\xi}{\underset{=
}{\Omega}}\frac{e^{\xi}}{\xi\left(1+\frac{\sqrt{A}}{2\xi}\right)}\right)=\frac{A_{\pm}}{\sqrt{A}}\left(e^{\sqrt{A}/2}-e^{-\sqrt{A}/2}\right)
=\frac{2A_{\pm}}{\sqrt{A}}\sinh\left(\frac{\sqrt{A}}{2}\right).
\end{align*} Therefore, we obtain
\begin{align}\label{XLHS}
\hat{X}(\boldsymbol{A})
=\left(\begin{array}{cc}
\cosh\left(\frac{\sqrt{A}}{2}\right)+ \frac{A_0}{\sqrt{A}}\sinh\left(\frac{\sqrt{A}}{2}\right)&\frac{2A_+}{\sqrt{A}}\sinh\left(\frac{\sqrt{A}}{2}\right)\\
-\frac{2A_-}{\sqrt{A}}\sinh\left(\frac{\sqrt{A}}{2}\right)&\cosh\left(\frac{\sqrt{A}}{2}\right)-\frac{A_0}{\sqrt{A}}\sinh\left(\frac{\sqrt{A}}{2}\right)
\end{array}\right)
\end{align}

Next, we consider Eq. \eqref{XDisent}. We have
\[
\exp\left(z\hat{K}^{\pm}\right)
=\hat{I}+z\hat{K}^{\pm}
\] using
\[
(\hat{K}^{\pm})^2=\left(\begin{array}{cc}
0&0\\
0&0
\end{array}\right).
\] Since $\hat{K}^0$ is diagonal, it is straightforward to obtain
\[
\exp\left(g\hat{K}^0\right)
=\left(\begin{array}{cc}
e^{g/2}&0\\
0&e^{-g/2}
\end{array}\right).
\] A direct matrix multiplication gives
\begin{align}\label{XRHS}
\hat{X}(\boldsymbol{A})
=\left(\begin{array}{cc}
1&f\\
0&1
\end{array}\right)\left(\begin{array}{cc}
e^{g/2}&0\\
0&e^{-g/2}
\end{array}\right)\left(\begin{array}{cc}
1&0\\
-h&1
\end{array}\right)=\left(\begin{array}{cc}
e^{g/2}-fe^{-g/2}h&fe^{-g/2}\\
-e^{-g/2}h&e^{-g/2}
\end{array}\right).
\end{align}

By comparing Eqs. \eqref{XLHS} and \eqref{XRHS} we obtain the equations
\begin{align}\label{galpha}
e^{-g/2}=\cosh\left(\frac{\sqrt{A}}{2}\right)- \frac{A_0}{\sqrt{A}}\sinh\left(\frac{\sqrt{A}}{2}\right),
\end{align}
\begin{align}\label{falpha}
fe^{-g/2}=\frac{2A_+}{\sqrt{A}}\sinh\left(\frac{\sqrt{A}}{2}\right),
\end{align} and
\begin{align}\label{halpha}
e^{-g/2}h
=\frac{2A_-}{\sqrt{A}}\sinh\left(\frac{\sqrt{A}}{2}\right).
\end{align}
We now show that our general result implies the factorization of the exponential operator involving the su(1,1) generators in Eq. \eqref{12}. Let us take $A_{\pm}=2i\beta Z$ and $A_0=2i\lambda Z$ so that $\sqrt{A}=2\Gamma Z$ and use Eqs. \eqref{galpha} and \eqref{falpha} or \eqref{halpha} (recall that here $A_+=A_-$) to obtain $g_0(Z)$ and $g_1(Z)$ in Eq. \eqref{12}.

Finally, we show how the matrix elements of the su(1,1) generators in the oscillator basis can be obtained using OMC. Recall that the su(1,1) Lie algebra admits the oscillator realization (see, e.g., \cite{82})
\[
\hat{K}^+\equiv \frac{\hat{a}^{\dagger 2}}{2},\quad
\hat{K}^0\equiv \frac{(\hat{n}+1/2)}{2},\quad {\rm and}\quad
\hat{K}^-\equiv \frac{\hat{a}^2}{2}\]
with $\hat{n}=\hat{a}^{\dagger}\hat{a}$. In what follows we consider only $\exp(h\hat{K}^-)$ with $\exp(f\hat{K}^+)$ being similar and $\exp(g\hat{K}^0)$ straightforward since $\hat{K}^0$ is diagonal. We have
\begin{align*}
\bra{m}\exp(h\hat{K}^-)\ket{n}&=\overset{\xi}{\underset{=
}{\Omega}}\exp\left(\frac{h\xi}{2}\right)\bra{m}\left(\hat{I}-\frac{\hat{a}^2}{\xi}\right)^{-1}\ket{n}\\
&=\overset{\xi}{\underset{=
}{\Omega}}\frac{\exp\left(\frac{h\xi}{2}\right)}{2}\left(\bra{m}\left(\hat{I}-\frac{\hat{a}}{\sqrt{\xi}}\right)^{-1}\ket{n}+\bra{m}\left(\hat{I}+\frac{\hat{a}}{\sqrt{\xi}}\right)^{-1}\ket{n}\right)\\
&=\frac{1}{2}\sum_{k\geq 0}\overset{\xi}{\underset{=
}{\Omega}}\exp\left(\frac{h\xi}{2}\right)\frac{1+(-1)^k}{\xi^{k/2}}
\bra{m}\hat{a}^k\ket{n}\\
&=\frac{1}{2}\sum_{k\geq 0}\overset{\xi}{\underset{=
}{\Omega}}\exp\left(\frac{h\xi}{2}\right)\frac{1+(-1)^k}{\xi^{k/2}}\sqrt{n(n-1)\cdots (n-k+1)}\delta_{m,n-k}\\
&=\sqrt{n(n-1)\cdots (m+1)}\overset{\xi}{\underset{=
}{\Omega}}\frac{\exp\left(\frac{h\xi}{2}\right)}{\xi^{(n-m)/2}}\\
&=\sqrt{\frac{n!}{m!}}\frac{\left(\frac{h}{2}\right)^{(n-m)/2}}{\left(\frac{n-m}{2}\right)!}
\end{align*} with $\delta$ meaning the Kronecker delta and $n\geq m$ with $n-m$ necessarily even in agreement with \cite[Eq.~(B.2)]{61}. Note that the first equality follows from Eq. \eqref{Omegaexp} and we used a partial fraction decomposition to obtain the second equality above. The aforementioned two restrictions on $m$ and $n$ can be rewritten in terms of two auxiliary functions, namely, a step function and the cosine to obtain $\mathscr{S}_{m,k}\left(Z\right)$.

\end{document}